\documentclass[aps,prl,reprint,raggedbottom,amsmath,amssymb]{revtex4-2}
\usepackage[utf8]{inputenc}
\usepackage{graphicx,tikz}
\usetikzlibrary{arrows.meta,decorations.pathreplacing}
\usepackage[hidelinks]{hyperref}
\allowdisplaybreaks

\begin{document}

\title{A new toric ALF gravitational instanton}
\author{Alfie Jones}
\affiliation{School of Mathematics and Maxwell Institute for Mathematical Sciences, University of Edinburgh, Edinburgh EH9 3FD, UK}

\begin{abstract}
   We construct a new explicit toric ALF gravitational instanton. 
   The instanton has Euler characteristic $\chi = 4$ and Hirzebruch signature $\tau = -1$. The topology at infinity is a lens space $L(3,1)$.
   The solution depends on one parameter and is constructed from the Belinski--Zakharov four-soliton solution. 
   We prove that the instanton is smooth and positive definite on and away from the axis.
   This is the first example of a non-Hermitian ALF (not AF) gravitational instanton.
\end{abstract}
\maketitle

\section{Introduction}

Gravitational instantons were introduced by Hawking in \cite{Hawking_1977}, by analogy with Yang--Mills instantons. Famous examples include the Euclidean Schwarzschild and Euclidean Kerr instantons \cite{Gibbons1979}. Originally, gravitational instantons were introduced to study Euclidean quantum gravity, where they are stationary phase points for the gravitational path integral.

A gravitational instanton $(M, g)$ is a complete Riemannian manifold satisfying the vacuum Einstein equations. Additionally, one often requires a curvature decay condition,
\(
\int_{M} | \operatorname{Riem} |^2 < \infty .
\)
Gravitational instantons exhibit a variety of asymptotic types, the most common of which are asymptotically locally Euclidean (ALE) and asymptotically locally flat (ALF). ALE instantons have quartic volume growth and approach $\mathbb R^4 / \Gamma$ for $\Gamma \subset SO(4)$ a finite subgroup. ALF instantons have cubic volume growth with boundary at infinity $S^3 / \Gamma$ or $S^1 \times S^2$. The latter case is an important subclass often called asymptotically flat (AF) and includes the Euclidean Kerr and Schwarzschild examples.

Initially, much work was done in understanding hyper-K\"ahler solutions, which necessarily have self-dual Riemann tensors. 
These solutions have been understood \cite{Kronheimer:1989zs, minerbe2009}, but much less is known outside of this case.

A natural conjecture is the Euclidean analogue of the no-hair theorem \cite{PhysRevD.22.1837}. However, it was proven false via explicit counterexample by Chen and Teo in 2011 \cite{ChenTeo_2011_AF}.
The Kerr and Chen--Teo solutions are both Hermitian \cite{aksteiner_2022}, toric, asymptotically flat gravitational instantons. Here, toric means that the instanton admits an effective isometric torus action with at least one fixed point. 

In the toric case, the Einstein equations reduce to an axisymmetric harmonic map equation from $\mathbb R^3 \setminus \{z\text{-axis}\}$ to the upper half-plane $\mathbb H^2 = SL(2, \mathbb R) / SO(2)$. 
The singular boundary data for this harmonic map problem is given by the rod structure, 
the fixed point data of the torus action \cite{Chen_2010}. 
The rod structure splits the axis into a sequence of intervals $I_j = (z_{j-1}, z_j)$ called ``rods''; these are intervals with $S^1$ isotropy. The endpoints $z_j$ of the rods are the fixed points of the full torus action. 
A vector generator for the isotropy $S^1$ on each rod, called the ``rod vector'', is expressed as an integer linear combination of the $2\pi$-periodic torus generators. Given appropriate rod data and asymptotics, a unique solution to the harmonic map problem always exists \cite{Kunduri_2021, KL26} and is smooth away from the axis, but this does not guarantee a smooth extension of the metric to the axis. In general one has conical singularities over each rod.

Toric, Hermitian non-K\"ahler instantons were classified in the ALF case by Biquard and Gauduchon \cite{Biquard_2022} and in the ALE case by Araneda and Lucietti \cite{araneda2025torichermitianalegravitational}. In each case a finite list was discovered. The question of if there exists algebraically general toric gravitational instantons was answered by Li and Sun \cite{li_2025}, who proved the existence of an infinite family of AF instantons. This list, indexed by $n$ where $n+1$ is the number of fixed points, starts with Kerr ($n=1$) and Chen--Teo ($n=2$) and assigns a new AF gravitational instanton for each $n \geq 3$. The method of Li and Sun does not provide explicit metrics. By varying the harmonic map parameters, they were able to use a topological argument to prove that each cone angle passes through $2\pi$ simultaneously, thereby removing all conical singularities and allowing the metric to extend smoothly to the axis.

The next explicit Li--Sun instanton was found by Teo \cite{teo2026newasymptoticallyflatgravitational} as a member of the Euclidean double Kerr--NUT family, it is the first explicit non-Hermitian AF instanton. 
It was asked \cite[Question 8.10]{li_2025} if there exists an algebraically general ALF gravitational instanton. In this paper we will explicitly construct the first example of such an ALF (not AF) instanton. Our solution has Euler characteristic $\chi = 4$ and Hirzebruch signature $\tau = -1$ and asymptotic topology $L(3,1)$. We start from the Euclidean double Kerr--NUT metric and explicitly solve the conditions for absence of conical singularities exactly. We also introduce coordinates where all metric components become rational, allowing for a simple proof of positivity and smoothness away from the axis.

During the preparation of this paper, the author became aware of independent work \cite{AnderssonDahlNilssonGeroch} by Andersson, Dahl and Nilsson, who construct the same instanton as part of a broader study of toric gravitational instantons. Furthermore, Dunajski, Mason and Tod have recently constructed new ALF gravitational instantons with $5$ fixed points via a classification using twistor methods \cite{dunajski-2026}.

\section{Construction and rod structure}

The starting point for this solution is the so-called Euclidean double Kerr--NUT metric. 
This can be constructed using the Belinski--Zakharov method \cite{Belinski_1978,Belinski_2001,Chen_2016} with a $4$ soliton transformation starting from a flat Euclidean seed.
This metric was recently used by Teo in \cite{teo2026newasymptoticallyflatgravitational} to find a new asymptotically flat (AF) solution.
The metric is written in the canonical Weyl--Papapetrou form
\begin{equation}\label{eq:bz-metric}
\begin{aligned}
g={}&\frac{\mu_1\mu_2\mu_3\mu_4 F}{H}
(\mathrm d\psi+\omega\,\mathrm d\phi)^2
+\frac{\rho^2H}{\mu_1\mu_2\mu_3\mu_4 F}\,\mathrm d\phi^2\\
&+\frac{k_0H}{(\mu_1\mu_2\mu_3\mu_4)^3R_{11}R_{22}R_{33}R_{44}}
(\mathrm d\rho^2+\mathrm dz^2),
\end{aligned}
\end{equation}
where
\[
\begin{aligned}
\mu_i &= \sqrt{\rho^2+(z-z_i)^2}-(z-z_i),\quad i=1,\ldots,4,\\
R_{ij} &= \rho^2+\mu_i\mu_j.
\end{aligned}
\]
The metric functions $F, H, \omega$ and the $2\pi$ periodic rod vectors $\ell_1, \cdots, \ell_5$ are given in Appendix~\ref{app:bz4}. 
The Euclidean double Kerr--NUT metric is a locally Ricci-flat toric metric whose rod vectors depend on 
continuous parameters $(k_0, C_1, C_2, C_3, C_4)$ and positions $(z_1, z_2, z_3, z_4)$ with $z_1< z_2 < z_3 < z_4$ which correspond to the $z$-positions of the four fixed points in our solution.

Another simplifying assumption we make is a reflection-symmetric ansatz on the rod structure:
\begin{gather*}
    z_1=-\kappa^{2},\quad z_2=-\kappa^{2}\lambda,\quad z_3=\kappa^{2}\lambda,\quad z_4=\kappa^{2}, \label{ansatz1}\\
    C_1C_4=1,\quad C_2C_3=1. \label{ansatz2}
\end{gather*}
We enforce $\kappa^{2}>0$ and $0<\lambda<1$ which preserves the ordering of our fixed points. Note that we have also fixed the $z$ origin.

\begin{figure*}[t]
\centering
\begin{tikzpicture}[x=1.2cm, y=1cm, font=\small, line cap=round]
    \draw[thick, -{Latex[length=2mm]}]
        (0,0) -- (10,0);

    \node[above=6pt] at (1,0) {$(1,0)$};
    \node[above=6pt] at (3,0) {$(0,1)$};
    \node[above=6pt] at (5,0) {$(1,-1)$};
    \node[above=6pt] at (7,0) {$(-1,2)$};
    \node[above=6pt] at (9,0) {{$(2,-3)$}};

    \foreach \x in {2,4,6,8} {
        \fill (\x,0) circle (1.7pt);
    }
    \node[below=5pt] at (2,0) {$z_1=-\kappa^{2}$};
    \node[below=5pt] at (4,0) {$z_2=-\kappa^{2}\lambda$};
    \node[below=5pt] at (6,0) {$z_3=\kappa^{2}\lambda$};
    \node[below=5pt] at (8,0) {$z_4=\kappa^{2}$};
\end{tikzpicture}
\caption{The rod structure of the new ALF gravitational instanton.}
\label{fig:double-taub-bolt-rods}
\end{figure*}
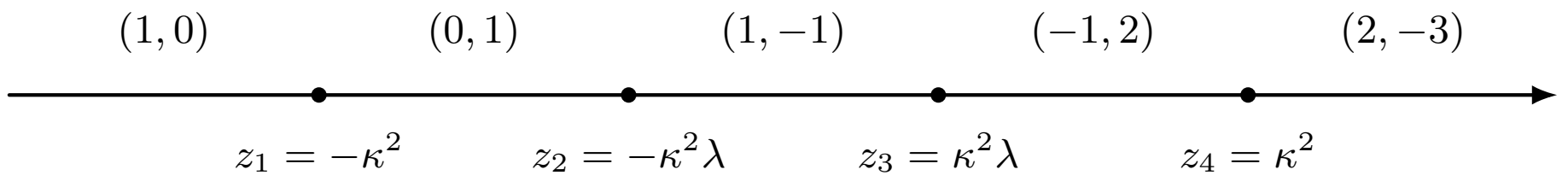

Our desired rod structure is listed in Figure \ref{fig:double-taub-bolt-rods}. 
To achieve this rod structure we impose the relations on our rod vectors
\begin{equation} \label{eq:rod}
\begin{aligned}
    \ell_1 &= \ell_2+\ell_3,  \\
    \ell_2 &= \ell_3+\ell_4, \\
    \ell_3 &= \ell_4+\ell_5. 
\end{aligned}
\end{equation}
These relations make it tempting to describe this instanton as a triple Taub-Bolt, since the $3$ consecutive rod vectors for Taub-Bolt satisfy \eqref{eq:rod}.
Using the rod vectors in Appendix~\ref{app:bz4}, the equations \eqref{eq:rod}
impose polynomial constraints on our parameters. These equations are homogeneous in $k_0$ and $\kappa$, 
hence we may consider them as equations in the variables $\lambda, C_1, C_2$.
Remarkably these equations have simple exact solutions. The one we are concerned with is:
\begin{equation}\label{eq:exact-sol}
\begin{aligned}
    C_1&=-3+\sqrt{2}-\sqrt{\frac{5}{2}}+\sqrt{5},\\
    C_2&=-2+\sqrt{2}+\sqrt{\frac{5}{2}}-\sqrt{5},\\
    \lambda&=\frac{7-2\sqrt{10}}{3}.
\end{aligned}
\end{equation}
It is simpler to work with a constant $\sigma$ related to $C_1, C_2$ in the following way:
\[
    C_1=\frac{\sigma-1}{\sigma+1},\qquad
    C_2=\frac{3\sigma+1}{3\sigma-1}.
\]
This has the exact value:
\[
    \sigma=\frac{-27+4\sqrt{2}+20\sqrt{5}-6\sqrt{10}}{123}>0.
\]
It is worth noting that equations \eqref{eq:rod} have $2$ valid solutions with $0 < \lambda < 1$, however
only \eqref{eq:exact-sol} leads to $H >0$ and therefore a valid instanton.
After imposing these equations we are left with $2$ parameters, $\kappa$ and $k_0$. 
However, the metric is invariant under the simultaneous substitutions
\begin{equation}\label{eq:scaling-symmetry}
\begin{gathered}
\kappa\rightarrow c^{1/2}\kappa,\qquad
k_0\rightarrow k_0/c^6,\\[3pt]
\rho\rightarrow c\rho,\quad z\rightarrow cz,\quad
\phi\rightarrow\phi/c,
\end{gathered}
\end{equation}
for any $c > 0$. This leaves us with a one parameter family of solutions.

\subsection{Topology}

The topology of a toric gravitational instanton is fixed by the rod structure \cite{Nilsson_2024}. In our case one can read off the topology by considering its toric-compactification; adding a bolt at infinity with $(1,-1)$ rod gives the rod structure of $\mathbb {CP}^2 \# 3\overline{\mathbb{CP}}^2$. Therefore our manifold has the topology of $(\mathbb{CP}^2 \# 3 \overline{\mathbb{CP}}^2) \setminus S^2$. This manifold has an Euler characteristic of $\chi = 4$ and Hirzebruch signature of $\tau = -1$ \cite{Gibbons1979}. 
In particular this allows us to demonstrate that our metric is non-Hermitian; Toric, hyper-K\"ahler ALF gravitational instantons are necessarily multi-Taub-NUT \cite{minerbe2009, KL26}. 
The unique such instanton with $\chi = 4$ is the $4$ centred Taub-NUT instanton \cite{gibbons-hawking-multi-center}, which has signature $\tau = \pm 3$. 
In \cite{Biquard_2022} Biquard and Gauduchon classified the toric ALF gravitational instantons which are Hermitian but not K\"ahler. All such instantons have Euler characteristic $\chi \leq 3$.
This proves that our metric is non-Hermitian.

\subsection{Asymptotic expansion}
Introduce polar coordinates on the Weyl half-plane by
\[
    \rho=R\sin\theta,
    \qquad
    z=R\cos\theta.
\]
We may also define functions $V, f$ as follows:
\[
\begin{aligned}
V&=g_{\psi\psi}=\frac{\mu_1\mu_2\mu_3\mu_4 F}{H},\\
f&=g_{\rho\rho}=g_{zz}
=\frac{k_0H}{(\mu_1\mu_2\mu_3\mu_4)^3R_{11}R_{22}R_{33}R_{44}}.
\end{aligned}
\]
Expanding these metric functions as $R\to \infty$ gives the asymptotics:
\[
\begin{aligned}
V & = 1 - 2m_0 \frac{ \kappa^{2}}{R} + O(R^{-2}),\\
f & = \alpha_0 k_0 \kappa^{8} \left( 1 + 2m_0 \frac{\kappa^{2}}{R} \right) + O(R^{-2}),\\
\omega & = N_0 \kappa^{2} \cos \theta + O(R^{-2}),
\end{aligned}
\]
where we have defined the constants
\begin{align*}
    N_0&=\frac{4(\lambda^2-1)
    \left[C_1(1-C_2^2)-\lambda C_2(1-C_1^2)\right]}{
    (1-\lambda)^2(1-C_1C_2)^2
    -(1+\lambda)^2(C_1-C_2)^2},\\[4pt]
    m_0&=\frac{(1-\lambda^2)
    \left[(1-\lambda)(1-C_1^2C_2^2)+(1+\lambda)(C_1^2-C_2^2)\right]}{
    (1-\lambda)^2(1-C_1C_2)^2
    -(1+\lambda)^2(C_1-C_2)^2},\\[4pt]
    \alpha_0&=\frac{16\bigl[(1-\lambda)^2(1-C_1C_2)^2
    -(1+\lambda)^2(C_1-C_2)^2\bigr]^2}{C_1^2C_2^2}.
\end{align*}
In order to match the asymptotic conventions found in \cite{KL26}, we require that $f \to 1$ as $R \to \infty$. We may achieve this by using the scaling symmetry \eqref{eq:scaling-symmetry} to fix
\begin{equation} \label{k0}
k_0=(\alpha_0\kappa^8)^{-1}.
\end{equation}
We will adopt this convention from now on. 
With these conventions the metric is ALF according to the definitions found in \cite{KL26}, with mass $m = \kappa^2 m_0$, angular momentum $j = 0$ and NUT charge $N=N_0 \kappa^2$.

\subsection{End Topology}
\label{sec:normalized-rods}
In the basis $\partial_\psi, \partial_\phi$ the rod vectors dramatically simplify to:
\begin{equation}\label{rods}
\begin{alignedat}{2}
    \ell_1 &= (N_0\kappa^2,1), \qquad
    &\ell_2 &= \left(N_0\kappa^2,\frac13\right),\\
    \ell_3 &= \left(0,\frac23\right), \qquad
    &\ell_4 &= \left(N_0\kappa^2,-\frac13\right),\\
    \ell_5 &= (-N_0\kappa^2,1).
\end{alignedat}
\end{equation}
One may check that equations \eqref{eq:rod} hold.  
The generators of the torus action can be taken to be $\ell_1, \ell_2$.
These rod vectors are all normalised such that their conical singularities are absent if their orbits are $2\pi$ periodic.
We wish to determine the periodicity of the coordinates $(\psi, \phi)$.
Write $\ell_1 = \partial_\chi$ and $\ell_2 = \partial_\eta$, for $\chi, \eta$ independently $2\pi$ periodic angles.
We may express $\psi, \phi$ in terms of $\chi, \eta$ as follows:
\[
    \psi=N_0 \kappa^{2}(\chi+\eta),
    \qquad
    \phi=\chi+\frac13\eta.
\]
We may therefore deduce that the coordinate identifications of $\psi, \phi$ are given as:
\[
\begin{gathered}
    (\psi,\phi)\sim(\psi+2\pi N_0 \kappa^{2},\phi+2\pi),
    \\
    (\psi,\phi)\sim
    \left(\psi+2\pi N_0 \kappa^{2},\phi+\frac{2\pi}{3}\right).
\end{gathered}
\]

The topology of a large-$R$ hypersurface is determined by the rod vectors for the semi-infinite rods
\[
\ell_1=\partial_\chi,\qquad
\ell_5=2\partial_\chi-3\partial_\eta.
\]
Writing $\ell_1 = \partial_{\varphi_2}$ and $\ell_5 = \partial_{\varphi_1}$ we deduce that
\[
\varphi_1=-\frac{\eta}{3},\qquad
\varphi_2=\chi+\frac{2\eta}{3},
\]
giving us the identifications
\[
\begin{gathered}
(\varphi_1,\varphi_2)\sim(\varphi_1,\varphi_2+2\pi),\\
(\varphi_1,\varphi_2)\sim
\left(\varphi_1+\frac{2\pi}{3},\varphi_2+\frac{2\pi}{3}\right).
\end{gathered}
\]
This identifies the topology at infinity as the lens space $L(3,1)$ (up to orientation).

\section{New Coordinates}

\label{sec:rational-coordinates}

The BZ metric explicitly depends on the square roots $$r_i = \sqrt{\rho^2 + (z-z_i)^2}.$$
In Appendix \ref{app:coordinate-domain} we construct coordinates $x,y$ away from the axis $\rho > 0$ 
which explicitly make all metric functions rational functions. 
These coordinates can be fully extended to the axis $\rho = 0$ by introducing $t = y / (1-x)$ near rod $5$.

To write the metric explicitly, introduce $x,y$ on the open rectangle:
\begin{equation}\label{eq:rational-domain}
\lambda<x<1,\qquad 0<y<1.
\end{equation}
We define the auxiliary functions $p, q$ which are the prolate coordinates adapted to outer two fixed points.
\begin{equation}\label{eq:rational-p}
p=\frac{1-x^2+(x^2-\lambda^2)y(2-y)}
{1-x^2+(x^2-\lambda^2)y^2},
\end{equation}
\begin{equation}\label{eq:rational-q}
q=\frac{x\bigl[(1-x^2)(1-2y)-(x^2-\lambda^2)y^2\bigr]}
{1-x^2+(x^2-\lambda^2)y^2}.
\end{equation}
These are invertible on the open rectangle \eqref{eq:rational-domain}, and the ranges of $p$, $q$ are given as $p > 1$ and $|q| < 1$.
In terms of these coordinates the original Weyl--Papapetrou coordinates may be expressed as
\begin{equation}\label{eq:rational-weyl-map}
\rho^2=\kappa^4(p^2-1)(1-q^2),\qquad
z=\kappa^2pq.
\end{equation}
Using $\tau=\psi/\kappa^2$, and the $x, y$ coordinates the metric becomes

\begin{widetext}
\nopagebreak[4]
\begin{equation}
g= \kappa^{4} \left[ \frac{\hat F}{\hat H}
\left(d\tau+\frac{\hat W}{\hat F}\,d\phi\right)^2
+\frac{(p^2 - 1)(1-q^2) \hat H}{ \hat F}\,d\phi^2
+\frac{\hat H}{144}\,\gamma \right],
\end{equation}
where in $x,y$ coordinates, $\gamma$ is given as:
\begin{equation}\label{eq:rational-orbit-metric}
\begin{aligned}
\gamma={}&
\frac{dx^2}{(1-x^2)(x^2-\lambda^2)} +
\frac{
\left([1-x^2+(x^2-\lambda^2)y^2]\,dy
+2xy(1-y)^2\,dx\right)^2
}{
y(1-y)[1-x^2+(x^2-\lambda^2)y]
[1-\{(x-\lambda)y-x\}^2]
[1-\{(x+\lambda)y-x\}^2]
}.
\end{aligned}
\end{equation}
The metric functions $(\hat F, \hat H, \hat W)$ are given by the expressions:
\[
\begin{aligned}
A={}&
\frac{5x^2+3(\lambda+\lambda^{-1})x+5}{2}\,p^2
-\frac{5+6\lambda/x+5\lambda^2/x^2}{2}\,q^2
+\frac{1-\lambda^2}{4\lambda\sigma}
[\lambda(1+9\sigma^2)+3x(1+\sigma^2)]p,
\\[4pt]
B={}&
-2(1-x^2)p^2
+\frac{2(x^2-\lambda^2)}{x^2}\,q^2
+\frac{1-\lambda^2}{4\lambda\sigma}
[\lambda(1-9\sigma^2)+3x(1-\sigma^2)]p,
\\[4pt]
\hat H={}&A^2-B^2,
\\[4pt]
\hat F={}&
\frac{
9(1-\lambda^2)^2(x+3\lambda)^2(x+\lambda/3)^2
}{
4\lambda^2(1-x^2)(x^2-\lambda^2)
}\,(p^2-1)(1-q^2)
+\frac{16(1-x^2)(x^2-\lambda^2)}{x^2}\,p^2q^2.
\end{aligned}
\]
Finally,
\[
\begin{aligned}
\hat W=16q\Bigg[&
\frac{
3(1-\lambda^2)^2(x+3\lambda)(x+\lambda/3)
[\lambda(1-9\sigma^2)+3x(1-\sigma^2)]
}{
64\lambda^2\sigma(x^2-\lambda^2)x
}
(p^2-1)(x^2p^2-\lambda^2)
\\
&+
\frac{1-\lambda^2}{8\lambda\sigma}
(1-x^2)[\lambda(1+9\sigma^2)+3x(1+\sigma^2)]
p^2(p^2-1)
+
\frac{3(1-\lambda^2)(1-x^2)(x^2-\lambda^2)}
{4\lambda x}
p^2\left(p+\frac{1+\sigma^2}{2\sigma}\right)
\Bigg].
\end{aligned}
\]
\end{widetext}
Despite appearances, these coordinates represent a major simplification of the metric functions, in particular the form of $\hat F$ is much simpler than $F$ in Weyl--Papapetrou coordinates, given in the appendix \ref{app:bz4}. We note that the constructed coordinates are valid on the axis away from rod $5$, or equivalently for $z < z_4$. 
We may avoid this problem by using a symmetry to provide two charts:
note that under $z \to -z$, the formulas \eqref{eq:rational-p}, \eqref{eq:rational-q} give 
\[
(p,q,x,y)\longmapsto
\left(
p,-q,x,\frac{2}{p+1}-y
\right).
\]
Under these transformations the metric functions transform as 
\[ (\hat F, \hat H, \hat W,\gamma) \to (\hat F, \hat H, - \hat W,\gamma) \]
Therefore, the full metric is invariant under 
\[ 
(\rho, z, \tau, \phi) \to (\rho, -z, \tau, -\phi)
\]
which exchanges rods $1$ and $5$, and $2$ and $4$.

\subsection{Positivity of $\hat F$ and $\hat H$}

While it is clear that $\hat{F} > 0$, it is less clear for $\hat{H}$. This section is dedicated to proving $\hat H > 0$.
For the ranges $\lambda<x<1$ and $0<y<1$, the formula for $p$ gives
\[
    p-1=\frac{2(x^2-\lambda^2)y(1-y)}
    {1-x^2+(x^2-\lambda^2)y^2}>0.
\]
The same formulas also give
\[
    \frac{x^2-\lambda^2}{x^2}(x^2-q^2)
    =(1-x^2)(p^2-1)>0.
\]
This implies that, $p>1$ and $|q|<x$. 

A direct computation using the expressions for $A$ and $B$ in the metric gives
\[
 \begin{aligned}
A+B={}&\frac{9x^2+3(\lambda+\lambda^{-1})x+1}{2}p^2
\\&-\frac{(x+3\lambda)^2}{2x^2}q^2
+\frac{(1-\lambda^2)(\lambda+3x)}{2\lambda\sigma}p,\\
A-B={}&\frac{x^2+3(\lambda+\lambda^{-1})x+9}{2}p^2
\\&-\frac{9x^2+6\lambda x+\lambda^2}{2x^2}q^2
+\frac{3(1-\lambda^2)\sigma(3\lambda+x)}{2\lambda}p.
\end{aligned}
\]
Both linear terms are positive. Using $p>1$ and $|q|<x$ gives
\[
\begin{aligned}
2(A+B)&>8x^2+3(\lambda^{-1}-\lambda)x\\
&\quad+1-9\lambda^2>0,\\
2(A-B)&>-8x^2+3(\lambda^{-1}-\lambda)x\\
&\quad+9-\lambda^2>0,
\end{aligned}
\]
where we have used $0< \lambda < 1/3$ and $\sigma > 0$ which can be checked from their exact values \eqref{eq:exact-sol}.
Therefore, $\hat H = A^2 - B^2 > 0$. This is enough to conclude that the metric is smooth and positive definite everywhere away from the axis.

We turn our attention to the axis $\rho = 0$, we need to verify that $\hat H > 0$ and $\hat F > 0$ on rods $1,2,3$ as then the $z \to -z$ symmetry guarantees this for rods $4,5$.

On the axis, the inequalities for the prolate functions become $p \geq 1$ and $|q| \leq x$, with $\lambda \leq x \leq 1$.
The linear terms in $A\pm B$ remain strictly positive, so the same lower bounds give $\hat H>0$.
After cancelling the apparent singularities at $x=1$ and $x=\lambda$, the boundary limits of $\hat F$ are:
\[
\begin{aligned}
\left.\hat F\right|_{\text{rod }1}&=144(p^2-1)(p^2-\lambda^2)>0,\\
\left.\hat F\right|_{\text{rod }2}&=16(1-q^2)(q^2-\lambda^2)>0,\\
\left.\hat F\right|_{\text{rod }3}&=64(\lambda^2-q^2)(1-q^2)>0.
\end{aligned}
\]
The coordinate ranges on each rod are given in Appendix~\ref{app:coordinate-domain}.
Note that $\hat F$ vanishes at the fixed points, but these inequalities hold on the interiors of the rods.

\section{Discussion}

We have constructed a smooth, toric, ALF gravitational instanton with $4$ fixed points which is non-Hermitian. The regularity conditions are solved exactly, and we found rational coordinates in which the metric functions all become rational and simplify.

It is interesting to note that the Killing field with finite norm at infinity has closed orbits. In our coordinates, this vector is $\partial_\psi$.
The generator of the corresponding $S^1$, with period $2\pi$, is therefore $u = 2N_0\kappa^2 \partial_\psi=-\ell_1+3\ell_2$.
Nevertheless, this solution does not satisfy the hypothesis of the extension of the Mars-Simon theorem \cite{MARS1999211} found in \cite[Proposition B.1]{li_2025}. This states that a toric ALF (not AF) instanton for which the Killing field with bounded norm generates an $S^1$ action with no discrete isotropy is either Type I (up to orientation) or Taub-Bolt. In our case, let $p$ lie on the first rod.
The action of $u$ is therefore: $\exp(s u) \cdot p = \exp(3 s \ell_2) \cdot p$
and hence $\operatorname{Stab}(p) \cong \mathbb Z_3$.
This is how our gravitational instanton evades the aforementioned uniqueness theorem.

\begin{acknowledgments}
I would like to thank James Lucietti and Bernardo Araneda for many useful discussions and Maciej Dunajski for useful comments. I am grateful to Lars Andersson, Mattias Dahl and Gustav Nilsson for sharing their work before publication.

The author's work was supported by the UKRI Centre for Doctoral Training in Algebra, Geometry and Quantum Fields (AGQ), under grant EP/Y035232/1.
\end{acknowledgments}

\bibliographystyle{apsrev4-2}
\bibliography{ALFrefs}

\onecolumngrid
\section*{End Matter}
\appendix
\setcounter{secnumdepth}{1}
\section{Full BZ4 metric}
\label{app:bz4}

For completeness, we give the general metric functions for \eqref{eq:bz-metric} and its rod
vectors~\cite{Chen_2016,teo2026newasymptoticallyflatgravitational}.
For fixed points $z_1<z_2<z_3<z_4$, set $\mu_{ij}=\mu_i-\mu_j, z_{ij}=z_i-z_j$.
\begingroup
\small
\begin{align*}
H&=\big[\rho^2\mu_{12}\mu_{13}\mu_{14}\mu_{23}\mu_{24}\mu_{34}(\rho^4+C_1C_2C_3C_4\mu_1\mu_2\mu_3\mu_4)
+R_{13}R_{14}R_{23}R_{24}\mu_{12}\mu_{34}(C_1C_2\mu_1\mu_2+C_3C_4\mu_3\mu_4)\\*
&\quad+R_{12}R_{13}R_{24}R_{34}\mu_{14}\mu_{23}(C_1C_4\mu_1\mu_4+C_2C_3\mu_2\mu_3)-R_{12}R_{14}R_{23}R_{34}\mu_{13}\mu_{24}(C_1C_3\mu_1\mu_3+C_2C_4\mu_2\mu_4)\big]^2\\*
&\quad
-\big[R_{12}R_{13}R_{14}\mu_{23}\mu_{24}\mu_{34}(\rho^2C_1\mu_1-C_2C_3C_4\mu_2\mu_3\mu_4)-R_{12}R_{23}R_{24}\mu_{13}\mu_{14}\mu_{34}(\rho^2C_2\mu_2-C_1C_3C_4\mu_1\mu_3\mu_4)\\*
&\quad+R_{13}R_{23}R_{34}\mu_{12}\mu_{14}\mu_{24}(\rho^2C_3\mu_3-C_1C_2C_4\mu_1\mu_2\mu_4)-R_{14}R_{24}R_{34}\mu_{12}\mu_{13}\mu_{23}(\rho^2C_4\mu_4-C_1C_2C_3\mu_1\mu_2\mu_3)\big]^2,\\*[3pt]
F&=\big[\rho^4(1+C_1C_2C_3C_4)\mu_{12}\mu_{13}\mu_{14}\mu_{23}\mu_{24}\mu_{34}
-(C_1C_2+C_3C_4)R_{13}R_{14}R_{23}R_{24}\mu_{12}\mu_{34}\\*
&\quad+(C_1C_3+C_2C_4)R_{12}R_{14}R_{23}R_{34}\mu_{13}\mu_{24}
-(C_1C_4+C_2C_3)R_{12}R_{13}R_{24}R_{34}\mu_{14}\mu_{23}\big]^2\\*
&\quad+\rho^2\big[(C_1+C_2C_3C_4)R_{12}R_{13}R_{14}\mu_{23}\mu_{24}\mu_{34}-(C_2+C_1C_3C_4)R_{12}R_{23}R_{24}\mu_{13}\mu_{14}\mu_{34}\\*
&\quad+(C_3+C_1C_2C_4)R_{13}R_{23}R_{34}\mu_{12}\mu_{14}\mu_{24}-(C_4+C_1C_2C_3)R_{14}R_{24}R_{34}\mu_{12}\mu_{13}\mu_{23}\big]^2,\\*[3pt]
\omega&=\frac{W}{\mu_1\mu_2\mu_3\mu_4F},\\*[3pt]
W&=2C_1C_2C_3R_{11}R_{14}R_{22}R_{24}R_{33}R_{34}\mu_{14}\mu_{24}\mu_{34}(\rho^2+C_4^2\mu_4^2)\\*
&\quad\times\big[R_{12}\mu_1\mu_3^2\mu_{12}(\rho^2-\mu_2^2)-R_{13}\mu_2^2\mu_3\mu_{13}(\rho^2-\mu_1^2)+R_{23}\mu_1^2\mu_2\mu_{23}(\rho^2-\mu_3^2)\big]\\*
&\quad
+2C_1C_2C_4R_{11}R_{13}R_{22}R_{23}R_{34}R_{44}\mu_{13}\mu_{23}\mu_{34}(\rho^2+C_3^2\mu_3^2)\\*
&\quad\times\big[R_{12}\mu_2\mu_4^2\mu_{12}(\rho^2-\mu_1^2)+R_{24}\mu_1^2\mu_4\mu_{24}(\rho^2-\mu_2^2)-R_{14}\mu_1\mu_2^2\mu_{14}(\rho^2-\mu_4^2)\big]\\*
&\quad
+2C_1C_3C_4R_{11}R_{12}R_{23}R_{24}R_{33}R_{44}\mu_{12}\mu_{23}\mu_{24}(\rho^2+C_2^2\mu_2^2)\\*
&\quad\times\big[R_{13}\mu_1\mu_4^2\mu_{13}(\rho^2-\mu_3^2)-R_{14}\mu_3^2\mu_4\mu_{14}(\rho^2-\mu_1^2)+R_{34}\mu_1^2\mu_3\mu_{34}(\rho^2-\mu_4^2)\big]\\*
&\quad
-2C_2C_3C_4R_{12}R_{13}R_{14}R_{22}R_{33}R_{44}\mu_{12}\mu_{13}\mu_{14}(\rho^2+C_1^2\mu_1^2)\\*
&\quad\times\big[R_{23}\mu_2\mu_4^2\mu_{23}(\rho^2-\mu_3^2)-R_{24}\mu_3^2\mu_4\mu_{24}(\rho^2-\mu_2^2)+R_{34}\mu_2^2\mu_3\mu_{34}(\rho^2-\mu_4^2)\big]\\*
&\quad-C_1R_{11}R_{12}R_{13}R_{14}\mu_{12}\mu_{13}\mu_{14}\big[
\rho^2\mu_{23}^2\mu_{24}^2\mu_{34}^2(\rho^6+C_2^2C_3^2C_4^2\mu_2^2\mu_3^2\mu_4^2)+R_{23}^2R_{24}^2\mu_{34}^2(\rho^2C_2^2\mu_2^2+C_3^2C_4^2\mu_3^2\mu_4^2)\\*
&\quad+R_{23}^2R_{34}^2\mu_{24}^2(\rho^2C_3^2\mu_3^2+C_2^2C_4^2\mu_2^2\mu_4^2)+R_{24}^2R_{34}^2\mu_{23}^2(\rho^2C_4^2\mu_4^2+C_2^2C_3^2\mu_2^2\mu_3^2)\big]\\*
&\quad+C_2R_{12}R_{22}R_{23}R_{24}\mu_{12}\mu_{23}\mu_{24}\big[
\rho^2\mu_{13}^2\mu_{14}^2\mu_{34}^2(\rho^6+C_1^2C_3^2C_4^2\mu_1^2\mu_3^2\mu_4^2)
+R_{13}^2R_{14}^2\mu_{34}^2(\rho^2C_1^2\mu_1^2+C_3^2C_4^2\mu_3^2\mu_4^2)\\*
&\quad+R_{13}^2R_{34}^2\mu_{14}^2(\rho^2C_3^2\mu_3^2+C_1^2C_4^2\mu_1^2\mu_4^2)+R_{14}^2R_{34}^2\mu_{13}^2(\rho^2C_4^2\mu_4^2+C_1^2C_3^2\mu_1^2\mu_3^2)\big]\\*
&\quad-C_3R_{13}R_{23}R_{33}R_{34}\mu_{13}\mu_{23}\mu_{34}\big[
\rho^2\mu_{12}^2\mu_{14}^2\mu_{24}^2(\rho^6+C_1^2C_2^2C_4^2\mu_1^2\mu_2^2\mu_4^2)+R_{12}^2R_{14}^2\mu_{24}^2(\rho^2C_1^2\mu_1^2+C_2^2C_4^2\mu_2^2\mu_4^2)\\*
&\quad+R_{12}^2R_{24}^2\mu_{14}^2(\rho^2C_2^2\mu_2^2+C_1^2C_4^2\mu_1^2\mu_4^2)+R_{14}^2R_{24}^2\mu_{12}^2(\rho^2C_4^2\mu_4^2+C_1^2C_2^2\mu_1^2\mu_2^2)\big]\\*
&\quad+C_4R_{14}R_{24}R_{34}R_{44}\mu_{14}\mu_{24}\mu_{34}\big[\rho^2\mu_{12}^2\mu_{13}^2\mu_{23}^2(\rho^6+C_1^2C_2^2C_3^2\mu_1^2\mu_2^2\mu_3^2)
+R_{12}^2R_{13}^2\mu_{23}^2(\rho^2C_1^2\mu_1^2+C_2^2C_3^2\mu_2^2\mu_3^2)\\*
&\quad+R_{12}^2R_{23}^2\mu_{13}^2(\rho^2C_2^2\mu_2^2+C_1^2C_3^2\mu_1^2\mu_3^2)+R_{13}^2R_{23}^2\mu_{12}^2(\rho^2C_3^2\mu_3^2+C_1^2C_2^2\mu_1^2\mu_2^2)\big]\,.
\end{align*}
\endgroup

\begingroup
\small
\begin{align*}
\ell_1&=\big({-}8\sqrt{k_0}(C_1C_2C_3z_{12}z_{13}z_{23}-C_1C_2C_4z_{12}z_{14}z_{24}+C_1C_3C_4z_{13}z_{14}z_{34}-C_2C_3C_4z_{23}z_{24}z_{34}),\\
&\qquad -4\sqrt{k_0}(C_1C_2z_{12}z_{34}-C_1C_3z_{13}z_{24}+C_1C_4z_{14}z_{23}+C_2C_3z_{14}z_{23}-C_2C_4z_{13}z_{24}+C_3C_4z_{12}z_{34})\big)\,,\\
\ell_2&=\big(8\sqrt{k_0}(C_2C_3z_{12}z_{13}z_{23}-C_2C_4z_{12}z_{14}z_{24}+C_3C_4z_{13}z_{14}z_{34}-C_1C_2C_3C_4z_{23}z_{24}z_{34}),\\
&\qquad 4\sqrt{k_0}(C_2z_{12}z_{34}-C_3z_{13}z_{24}+C_4z_{14}z_{23}+C_1C_2C_3z_{14}z_{23}-C_1C_2C_4z_{13}z_{24}+C_1C_3C_4z_{12}z_{34})\big)\,,\\
\ell_3&=\big(8\sqrt{k_0}(C_3z_{12}z_{13}z_{23}-C_4z_{12}z_{14}z_{24}-C_1C_3C_4z_{23}z_{24}z_{34}+C_2C_3C_4z_{13}z_{14}z_{34}),\\
&\qquad 4\sqrt{k_0}(z_{12}z_{34}+C_1C_3z_{14}z_{23}-C_1C_4z_{13}z_{24}-C_2C_3z_{13}z_{24}+C_2C_4z_{14}z_{23}+C_1C_2C_3C_4z_{12}z_{34})\big)\,,\\
\ell_4&=\big({-}8\sqrt{k_0}(z_{12}z_{13}z_{23}-C_1C_4z_{23}z_{24}z_{34}+C_2C_4z_{13}z_{14}z_{34}-C_3C_4z_{12}z_{14}z_{24}),\\
&\qquad -4\sqrt{k_0}(C_1z_{14}z_{23}-C_2z_{13}z_{24}+C_3z_{12}z_{34}+C_1C_2C_4z_{12}z_{34}-C_1C_3C_4z_{13}z_{24}+C_2C_3C_4z_{14}z_{23})\big)\,,\\
\ell_5&=\big(8\sqrt{k_0}(C_1z_{23}z_{24}z_{34}-C_2z_{13}z_{14}z_{34}+C_3z_{12}z_{14}z_{24}-C_4z_{12}z_{13}z_{23}),\\
&\qquad -4\sqrt{k_0}(C_1C_2z_{12}z_{34}-C_1C_3z_{13}z_{24}+C_1C_4z_{14}z_{23}+C_2C_3z_{14}z_{23}-C_2C_4z_{13}z_{24}+C_3C_4z_{12}z_{34})\big)\,.
\end{align*}
\endgroup

\twocolumngrid
\section{Construction of the rational coordinates}
\label{app:coordinate-domain}
The metric depends on the square roots $r_i$:
\[
r_i=\sqrt{\rho^2+(z-z_i)^2},\qquad i=1,2,3,4.
\]
We look for coordinates where all of these functions become rational expressions.
We first start with prolate coordinates \citep[Appendix A]{Chen_2016} for $r_1, r_4$.
These are defined as follows:
\begin{equation}\label{eq:rational-prolate}
\begin{aligned}
    p &= \frac{r_1+r_4}{2\kappa^2},\qquad q = \frac{r_1-r_4}{2\kappa^2},\\
    \rho^2 &= \kappa^4(p^2-1)(1-q^2),\qquad z = \kappa^2pq,\\
\end{aligned}
\end{equation}
with the ranges $p \ge 1$ and $-1\le q\le 1$.
We have the relation \[r_2^2 - r_3^2 = 4 \lambda \kappa^2 z = 4 \lambda \kappa^4 pq .\]
We define our $x$ coordinate as follows:
\[x = \frac{r_2 + r_3}{2\kappa^2p}.\] 
We have $r_2 + r_3 = 2 \kappa^2 p x$ and {$r_2 - r_3 = \frac{2\kappa^2 \lambda q}{x}$}.
The other relation for $r_2, r_3$ simplifies to:
\begin{equation}\label{eq:rational-conic}
(1-x^2)p^2 + \frac{x^2 - \lambda^2}{x^2}q^2=1-\lambda^2.
\end{equation}
This defines an ellipse, with known point $(p,q) = (-1, x)$.
If we parametrise this conic by lines through this point, we get coordinate $y$ defined as: 
\begin{equation}\label{eq:rational-line}
q=x(1-y(p+1)) .
\end{equation}
Solving, we get $p(x,y), q(x,y)$ rational as needed.
The explicit expressions are \eqref{eq:rational-p}, \eqref{eq:rational-q}.
It is worth noticing that we have completed our goal of rationalising all $r_i$, we have:
\begin{equation}\label{eq:rational-inner-distances}
\begin{aligned}
r_1&=\kappa^2(p+q),\\
r_2&=\kappa^2\left(xp+\frac{\lambda q}{x}\right),\\
r_3&=\kappa^2\left(xp-\frac{\lambda q}{x}\right),\\
r_4&=\kappa^2(p-q).
\end{aligned}
\end{equation}
Finally, consider the coordinate ranges for $\rho>0$ and $0<\lambda<1$.
We have $p>1$, $|q|<1$, and
\[
    x=\frac{r_2+r_3}{r_1+r_4},\qquad
    y=\frac{x-q}{x(p+1)}.
\]
To bound $x$, define the $\mathbb R^2$ vectors $v_i=(\rho,z-z_i)$, such that $\|v_i\|=r_i$, we write
$a=(1+\lambda)/2$, $b=(1-\lambda)/2$. Then
\[
\begin{aligned}
    v_2&=a v_1+b v_4,&
    v_3&=b v_1+a v_4,\\
    \lambda v_1&=a v_2-b v_3,&
    \lambda v_4&=a v_3-b v_2.
\end{aligned}
\]
Applying the triangle inequality to each row and summing gives
\[
    \lambda(r_1+r_4)<r_2+r_3<r_1+r_4.
\]
Similarly, the reverse triangle inequality gives
\[
    |r_2-r_3|<\|v_2-v_3\|=2\lambda \kappa^2 .
\]
Since $r_2-r_3=2\kappa^2\lambda q/x$, this yields $|q|<x$. Consequently,
\[
    0<y=\frac{x-q}{x(p+1)}<\frac{2}{p+1}<1.
\]
Thus the coordinate ranges are $\lambda<x<1$, $0<y<1$, with
$p>1$ and $|q|<x<1$ throughout the open domain.

To compute the values of $x,y$ on each rod we use the equations
$x = \frac{r_2 + r_3}{r_1+r_4}$ and $y = \frac{x-q}{x(p+1)}$

\begin{itemize}
    \item The leftmost rod has $z < -\kappa^2$. We have $(p,q) = (p, -1)$ for $p > 1$, or $(x,y) = (1, y)$ for $0 < y < 1$.
    \item The next rod has $-\kappa^2 < z < -\lambda \kappa^2$. We have $(p,q) = (1, q)$ for $-1 < q < -\lambda$, or $(x,y) = (x, 1)$ for $\lambda < x < 1$.
    \item The middle rod has $-\lambda \kappa^2 < z < \lambda \kappa^2$. We have $(p,q) = (1, q)$ for $-\lambda < q < \lambda$, or $(x,y) = (\lambda, y)$ for $0 < y < 1$.
    \item The next rod has $\lambda \kappa^2 < z < \kappa^2$. We have $(p,q) = (1, q)$ for $\lambda < q < 1$, or $(x,y) = (x, 0)$ for $\lambda < x < 1$.
    \item The rightmost rod has $z > \kappa^2$. We have $(p,q) = (p, 1)$ for $p > 1$, or $(x,y) = (1, 0)$. This is degenerate in $(x,y)$ coordinates.
\end{itemize}

Note that in these coordinates the rod $z > \kappa^2$ collapses to the corner $(x,y) = (1,0)$.

This degeneration at the rightmost rod can be resolved either by introducing the coordinate $t=y/(1-x)$, or by using the reflection symmetry $z\mapsto -z$ 
to obtain a second coordinate chart.
The main downside of the $t$ coordinate is that we lose the rectangular domain.

To cast the metric in these coordinates, it helps to first simplify the metric functions that show up in the Belinski--Zakharov metric.
We will begin by pulling out a common factor from the terms in $F$, $H$ and $W$. To do this it will be useful to have the identity
\begin{equation}\label{eq:rational-soliton-identity}
R_{ij}\mu_{ij}=2\mu_i\mu_jz_{ij}.
\end{equation}
To prove this it is enough to note that $$\mu_i^2=2(z_i-z)\mu_i+\rho^2.$$
The functions $F, H, W$ are then reduced as follows:
\[
\begin{aligned}
F&=\frac{2^{18}\lambda^2\sigma^4\kappa^{16}}
{(1-\sigma^2)^2(1-9\sigma^2)^2}
(\mu_1\mu_2\mu_3\mu_4)^3\hat F,\\
H&=\frac{2^{18}\lambda^2\sigma^4\kappa^{16}}
{(1-\sigma^2)^2(1-9\sigma^2)^2}
(\mu_1\mu_2\mu_3\mu_4)^4\hat H,\\
\omega&=\frac{W}{\mu_1\mu_2\mu_3\mu_4 F}
=\kappa^2\frac{\hat W}{\hat F},\\
W&=\frac{2^{18}\lambda^2\sigma^4\kappa^{18}}
{(1-\sigma^2)^2(1-9\sigma^2)^2}
(\mu_1\mu_2\mu_3\mu_4)^4\hat W.
\end{aligned}
\]
Additionally, it is helpful to have the results
\[
\begin{aligned}
\mu_1&=\kappa^2(p-1)(1-q),\\
\mu_2&=\kappa^2\left(p-\frac{\lambda}{x}\right)(x-q),\\
\mu_3&=\kappa^2\left(p+\frac{\lambda}{x}\right)(x-q),\\
\mu_4&=\kappa^2(p+1)(1-q).
\end{aligned}
\]
Finally, the orbit space metric takes a simple form. Using the identities:
\[
\begin{gathered}
R_{ii}=2\mu_i r_i, \quad 
\frac{2^{18}k_0\lambda^2\sigma^4\kappa^8}
{(1-\sigma^2)^2(1-9\sigma^2)^2}=\frac19.
\end{gathered}
\]
The latter of which follows from $k_0 \alpha_0 \kappa^8 = 1$. 
Gives the orbit space metric as:
\[
\begin{aligned}
\frac{k_0H\,(d\rho^2+dz^2)}
{(\mu_1\mu_2\mu_3\mu_4)^3R_{11}R_{22}R_{33}R_{44}} &=\frac{\kappa^8\hat H}{144}
\frac{d\rho^2+dz^2}{r_1r_2r_3r_4}\\
&=\frac{\kappa^4\hat H}{144}\,\gamma,
\end{aligned}
\]
where we define $\gamma$ as
\[
\gamma
=\frac{1}{x^2p^2-\lambda^2q^2/x^2}
\left(\frac{dp^2}{p^2-1}+\frac{dq^2}{1-q^2}\right).
\]
In $x,y$ coordinates, this becomes \eqref{eq:rational-orbit-metric}.

\end{document}